\documentclass[aip,reprint]{revtex4-1}

\usepackage{graphicx}
\usepackage{subfigure}
\usepackage{color}

\draft 

\begin{document}


\title{Shock Wave in Leader Discharge Observed Using Mach-Zehnder Interferometry} 



\author{Yingzhe Cui}
\thanks{Cui, Zhuang and Zhou contributed equally to this work.}
\affiliation{State Key Lab of Power System, Department of Electrical Engineering, Tsinghua University, Beijing 100084, China}
\author{Chijie Zhuang}
\thanks{Cui, Zhuang and Zhou contributed equally to this work.}
\affiliation{State Key Lab of Power System, Department of Electrical Engineering, Tsinghua University, Beijing 100084, China}
\author{Xuan Zhou}
\thanks{Cui, Zhuang and Zhou contributed equally to this work.}
\affiliation{State Key Lab of Power System, Department of Electrical Engineering, Tsinghua University, Beijing 100084, China}
\author{Zezhong Wang}
\affiliation{State Key Lab of Power System, Department of Electrical Engineering, Tsinghua University, Beijing 100084, China}
\author{She Chen}
\affiliation{College of Electrical and Information Engineering, Hunan University, Changsha 410082, China.}
\author{Rong Zeng}
\email[Corresponding author. e-mail: ]{zengrong@tsinghua.edu.cn.}
\affiliation{State Key Lab of Power System, Department of Electrical Engineering, Tsinghua University, Beijing 100084, China}

%


\date{\today}

\begin{abstract}
A leader is an electric discharge mechanism in long-air-gap discharges. In this work, we report the shock-wave phenomenon in an air-gap leader discharge observed using a Mach-Zehnder interferometer with a time resolution of several microseconds. The continuous temporal evolution of the shock wave and the plasma channel was recorded and reproduced with a thermo-hydrodynamic model based on the measured current. The wave propagated at nearly the speed of sound, and the simulation results for the shock wave front positions and the plasma channel radius showed good consistency with the experimental measurements. Detailed thermal parameters obtained through the simulation showed that continuous energy injection by the current results in a temporary over-pressure process in the plasma channel and produces the shock wave.
\end{abstract}


\maketitle 

A leader is an electric discharge mechanism in long air gap discharges \cite{silva2017}, e.g., lightning \cite{big2017,zhu2017} and flashover between the power conductors and a tower in a power system \cite{far2018,rakov2018}. This letter reports the shock wave phenomenon in a 1-m-long-air-gap leader discharge observed using a Mach-Zehnder interferometer and reveals the physics behind the shock wave using a theoretical model.

\begin{figure}
\includegraphics[width=0.45\textwidth]{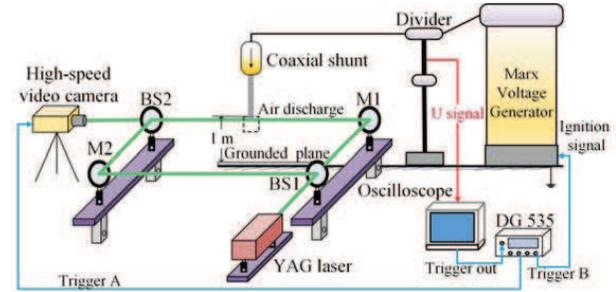}
\caption{\label{fig1}Schematic diagram of the experimental setup.}%
\end{figure}

A Mach-Zehnder interferometer consists of two reflector mirrors (M1 and M2), two beam splitters (BS1 and BS2), and a YAG laser (see FIG. \ref{fig1}). The laser beam emitted by the YAG laser is separated into two beams with equal intensities via B1: one is routed though the discharge channel, and the other beam is used as a reference. When discharge occurs, the particle density and the mass density in the channel may change accordingly, which causes the associated refractivity to change, meaning that interference fringes can be detected; by this way, the interferometer can record the dynamics in the discharge channel.

Air gap electrical discharge experiments under a positive impulse voltage with rise and half-wave times of 250 $\mu$s and 2500 $\mu$s, respectively, were conducted (see Fig. \ref{fig1}). A rod with a cone tip 2.5 cm in diameter and a length of 1 m was suspended above a well-grounded 2 m$\times$2 m aluminum sheet. A voltage impulse was applied to the 1-m rod-plane air gap and measured with a standard divider and an oscilloscope. A coaxial shunt was connected in series to the high-voltage electrode to measure the discharge current.

\begin{figure*}
\includegraphics[width=0.95\textwidth]{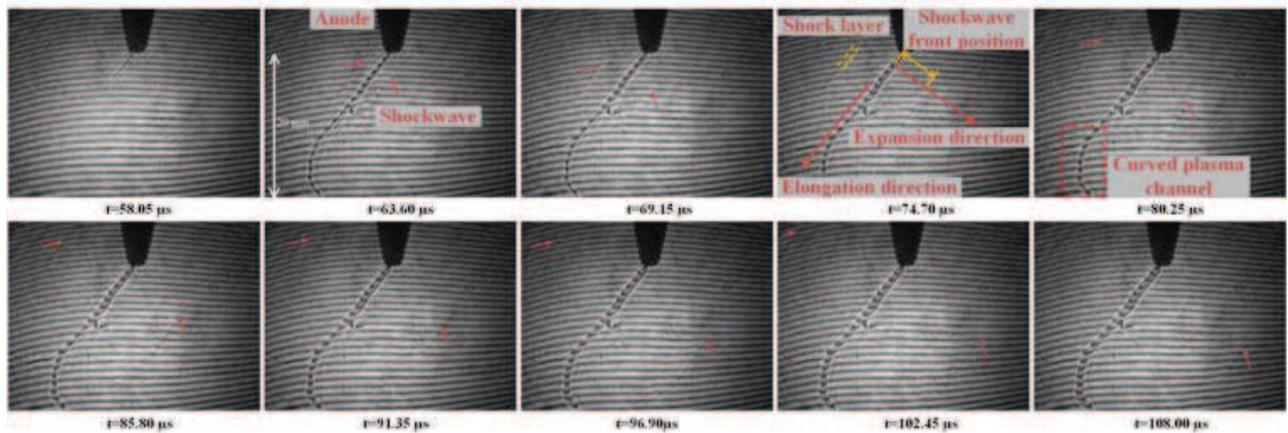}
\caption{\label{fig2}Typical interferometer images taken during the evolution of the shock wave and plasma channel in the 1-m rod-plane air gap. The exposure time and interval is 0.75 $\mu$s and 5.55 $\mu$s, respectively. The resolution is 192$\times$144 pixels for each image. The amplitude of the positive impulse voltage was 404 kV. Time=0 was taken as the interval the voltage pulse begins to rise.}%
\end{figure*}

The Mach-Zehnder interferometer was focused on the discharge area below the electrode tip to observe the discharge dynamics. The frequency-doubled YAG used was a continuous light source. The laser spot was 5 mm in diameter and expanded to 45 mm via a beam-expander in the YAG laser, which covered a considerable observation area near the tip of the rod. The distance between BS2 and M1 (also BS1 and M2) was set to 4 m to ensure sufficient insulation. A high-speed video camera (Plantom V12.1) was used to record the interferometer. The equipment was synchronized by a digital pulse/delay generator (DG535, Stanford Research). The DG535 received the trigger signal from the oscilloscope and triggered the high-speed video camera and voltage generator.

The experiments were conducted repeatedly and the corresponding interferometer images that show temporal and spatial evolution of the plasma channel were obtained. Interestingly, a shock wave appears in the streamer-to-leader transition and propagates as the plasma channel expands (see supplementary material). Some typical results are shown in FIG. 2. The plasma channel (specifically, the leader filament) appears at 58.05 $\mu$s, and then elongates at least 1.3$\times$10$^4$ m/s and expands at the same time. The plasma channel expands at 6.40 m/s, which is close to 6.70 m/s obtained under similar experimental conditions\cite{pop2015}.

By no more than 63.6 $\mu$s, a shock wave clearly appears. The shock layer was approximately 1.42 mm thick. From 63.6 to 108 $\mu$s, the shock wave front moves from 1.95 to 15.92 mm away from the center axis with an estimated velocity of 319 m/s, which is comparable to the speed of sound (typically 340 m/s). In addition, the shock wave produced by the curved plasma channel (see FIG. \ref{fig2}) also exists although is not very clear.

\begin{figure}
\includegraphics[width=0.45\textwidth]{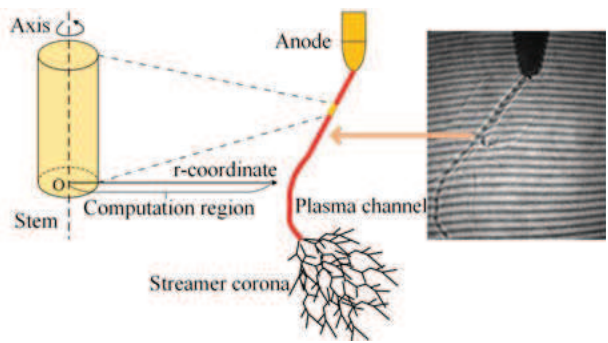}
\caption{\label{fig3}Schematic diagram of the model and computation region.}%
\end{figure}

As schematically represented in FIG. \ref{fig3}, the stem and plasma channel are continuously heated by the current produced by the filaments within the streamer corona zone and become thermalized, which drives the propagation of the plasma channel. It could elongate at a speed of about 6$\times10^4$ m/s or even higher \cite{suzuki1971,marode1975}, which is greater than the speed of sound by two orders of magnitude. Therefore, the plasma channel and the shock wave develop at different time scales and the plasma channel is reasonably assumed to form immediately when the expansion of the shock wave is considered. So, stems in the plasma channel will produce shock waves simultaneously. Upon further consideration of the axis symmetry, a cylindrical shock wave is formed (see FIG. \ref{fig2}). Based on this assumption, the profile of the shock wave is consistent with the shape of the plasma channel, that is, the profile of shock wave bends when the channel bends, which is confirmed by our experiments (see FIG. S2 and S3, and the videos, in the supplementary material).

\begin{figure}
\includegraphics[width=0.45\textwidth,height=2.1in]{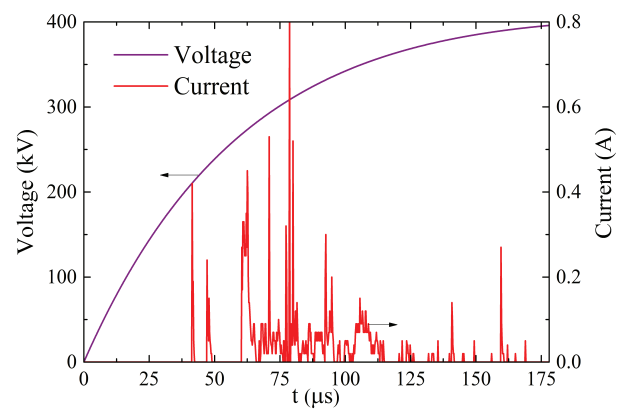}
\caption{\label{fig4}The measured discharge current and voltage.}%
\end{figure}

To study the shock wave dynamics observed in the experiment, a one-dimensional thermo-hydrodynamic model that assumes the stem is cylindrically symmetric (see FIG. \ref{fig3}) was employed\cite{popov2003,pasko2013}. The model takes into consideration gas dynamics that describe the air heating and radial expansion of a plasma channel{\cite{mikal2011, mikal2017}}, a kinetic scheme, energy exchange between charged and neutral particles, and a vibrational energy relaxation; as described in detail in {[\onlinecite{popov2003}] and [\onlinecite{pasko2013}]. The one-dimensional approximation is necessary due to the extensive physical and chemical processes with large time and space scales. Calculating using a fully 3-dimensional system is hardly affordable for personal computers.

The computation region (see FIG. \ref{fig3}) was set to 18 mm which is sufficiently large according to FIG. \ref{fig2} and was filled with a mixture of 79\% N$_2$ and 21\% O$_2$ at 300 K and 0.101 MPa for the initial conditions. Consistent with references [\onlinecite{popov2003}] and [\onlinecite{pasko2013}], only the electron and positive O$_2^+$ ion generated by ionization of oxygen molecules are reasonably assumed to exist initially in the channel, and their density follows a Gaussian distribution, i.e., the initial condition is $n_e=n_{\text{O}_2^+}=n_{e0}\exp{(-r^2/r_0^2)}$, where ${n_{e}}_0=2\times10^{20}$ m$^{-3}$ and $r_0=0.3$ mm\cite{popov2003,pasko2013}. A second-order MUSCL (Monotonic Upwind Scheme for Conservation Laws) and a second-order TVD Runge-Kutta scheme \cite{siam2001} were used for the space and temporal discretization, respectively. Notably, the measured discharge current shown in FIG. \ref{fig4}, not a hypothetical current\cite{pasko2013}, was used as the input in the simulation. Because only one plasma channel develops from the electrode according to FIG. \ref{fig2}, there was no need to assign the current for different stems.

FIG. \ref{fig5}(a) and \ref{fig5}(b) show the simulation results from the model. A shock wave is produced and expands as the plasma channel develops, and expands to the background atmosphere (see FIG. 5(a)). The experimental shock wave dynamics accompanying the changes in the refractive index reflect the changes in the mass density, consistent with results shown in FIG. \ref{fig5b}. To obtain the simulation results for the shock wave's front position and the plasma radius from the model, we take the maximum density positions as the shock wave front positions in FIG. \ref{fig5b}, and take the positions with relative density equal to 0.95 as the plasma boundary \cite{epsr2016}. The results from the simulation and the interferometer images agree well, with relative differences less than 10\% (see FIG. \ref{fig5c}).
\begin{figure*}[htbp]
\subfigure[]{
\label{fig5a} \includegraphics[width=0.45\textwidth, height=0.32\textwidth]{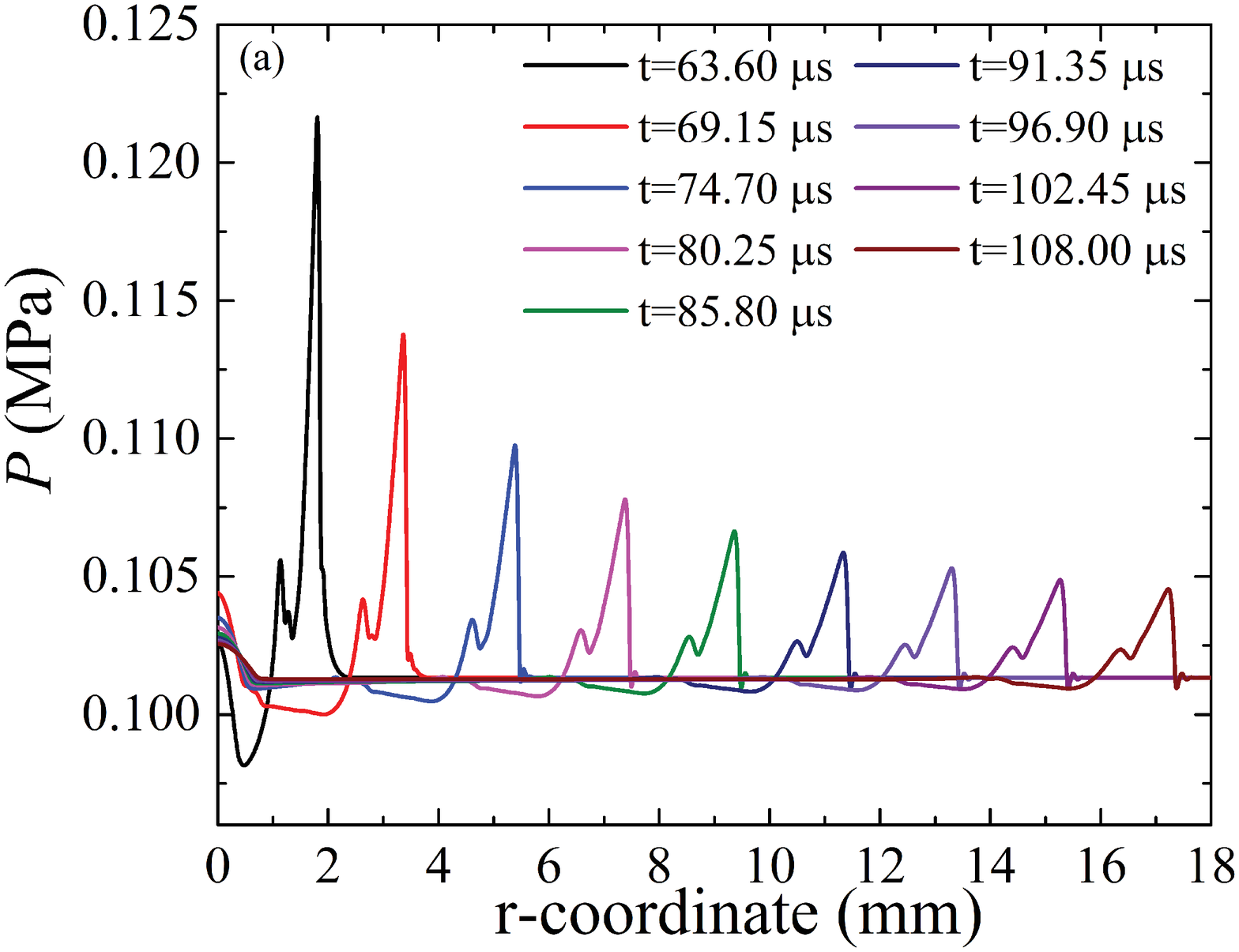}}
\subfigure[]{
\label{fig5b} \includegraphics[width=0.45\textwidth, height=0.32\textwidth]{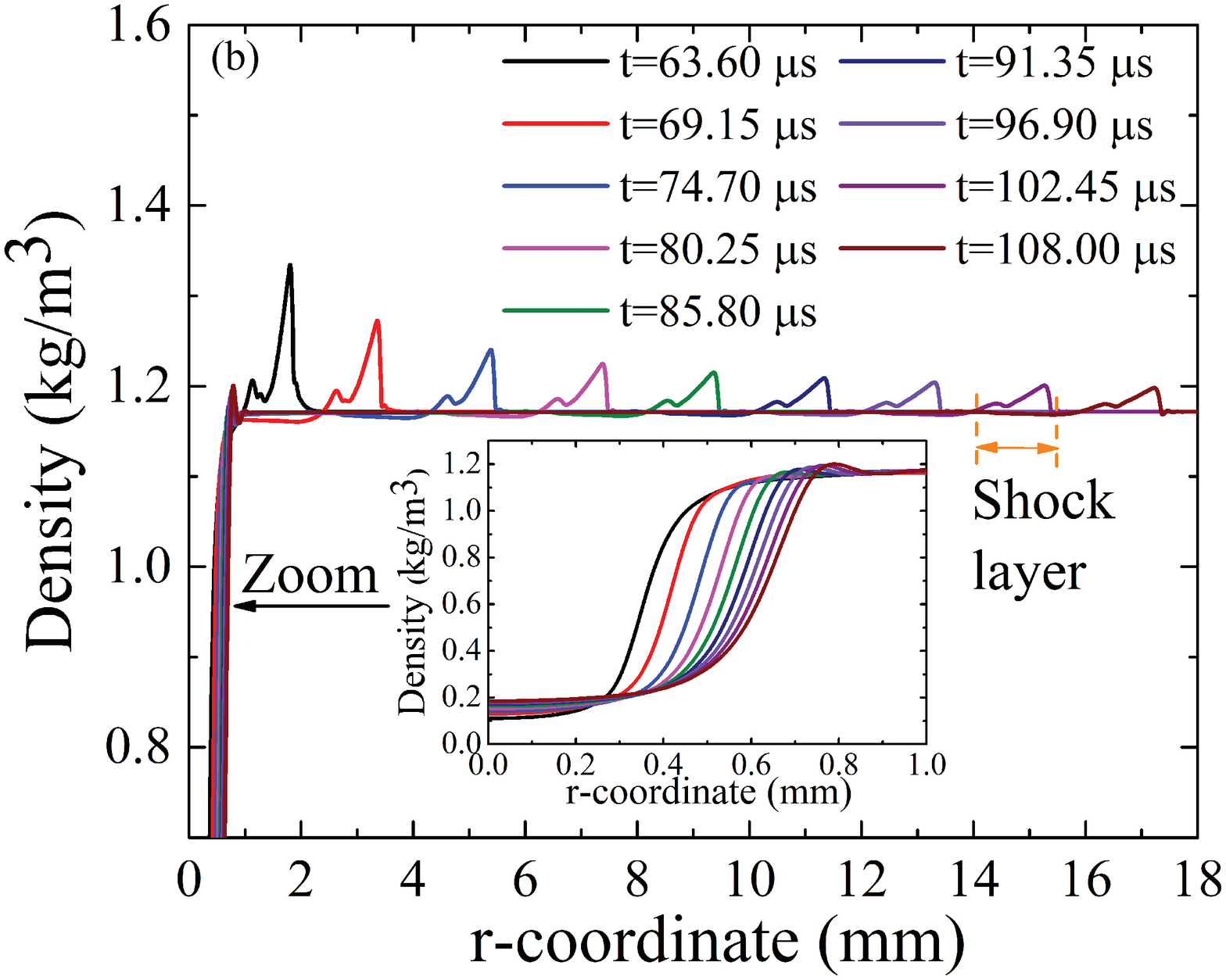}}
\subfigure[]{
\label{fig5c} \includegraphics[width=0.45\textwidth, height=0.32\textwidth]{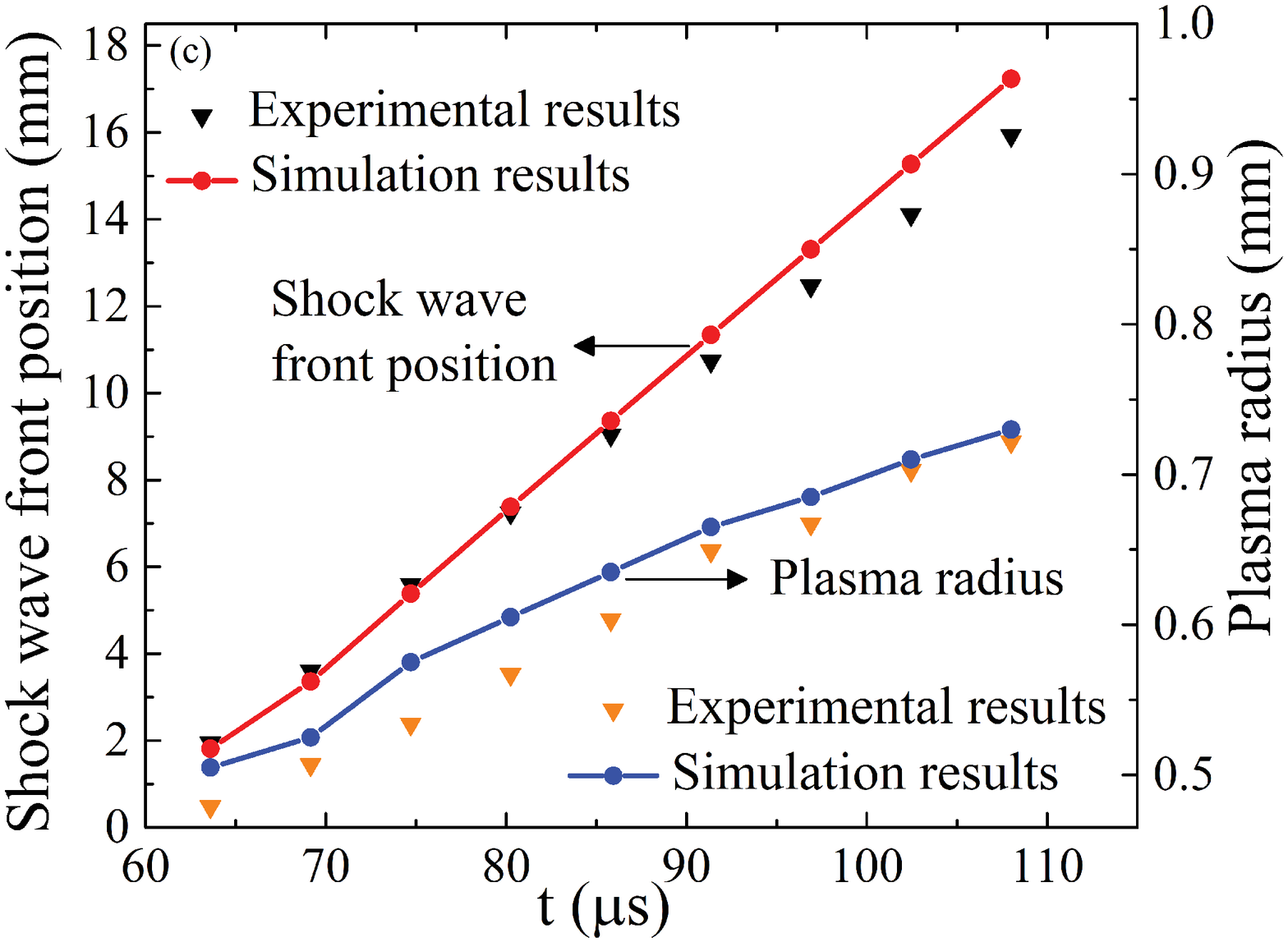}}
\caption{The radial gas dynamics of (a) pressure $P$, and (b) mass density obtained from the model. (c) Position of the shock wave front and plasma radius as a function of time, obtained from interferometer images and the model. The times are consistent with those in FIG. 2.}
\label{fig5}
\end{figure*}

The over-pressure in the plasma channel is the source of the shock wave. FIG. 6(a) shows a sharp increase in the temperature (4000 K) at the discharge axis due to the fast heating mechanism{\cite{popov2003, pasko2013}}. Before 61.3 $\mu$s, the fast heating power $Q_{\text{T}}$ shown in FIG. 6(b) mainly contributes to the expansion of the plasma channel which causes the convection loss. The corresponding electric field decreases to approximately 1 kV/cm, which agrees with the results of previous research{\cite{popov2003, pasko2013}}. Consequently, the plasma channel, with a rapid rise in pressure (see FIG. 6(a)), extrudes and pushes away the ambient gas. In FIG. 5(b), the mass swept by the shock wave forms a thin shock layer (1.4 mm, approaching the measured result in FIG. 2). The shock layer travels with the shock wave, leaving behind a lower-density region near the axis. The corresponding mass density is lower by approximately an order of magnitude than the normal ambient density (1.172 kg/m$^3$) (see FIG. 5(b)). After 61.3 $\mu$s, vibrational-translational relaxation $Q_{\text{VT}}$ becomes the dominant mechanism for air heating; however $Q_{\text{VT}}$ is less than the convection loss and thermal conduction loss. As a result, the temperature decreases and approaches approximately 2000 K (see FIG. 6(a)).
\begin{figure*}[htbp]
\subfigure[]{
\label{fig6a} \includegraphics[width=0.45\textwidth, height=0.32\textwidth]{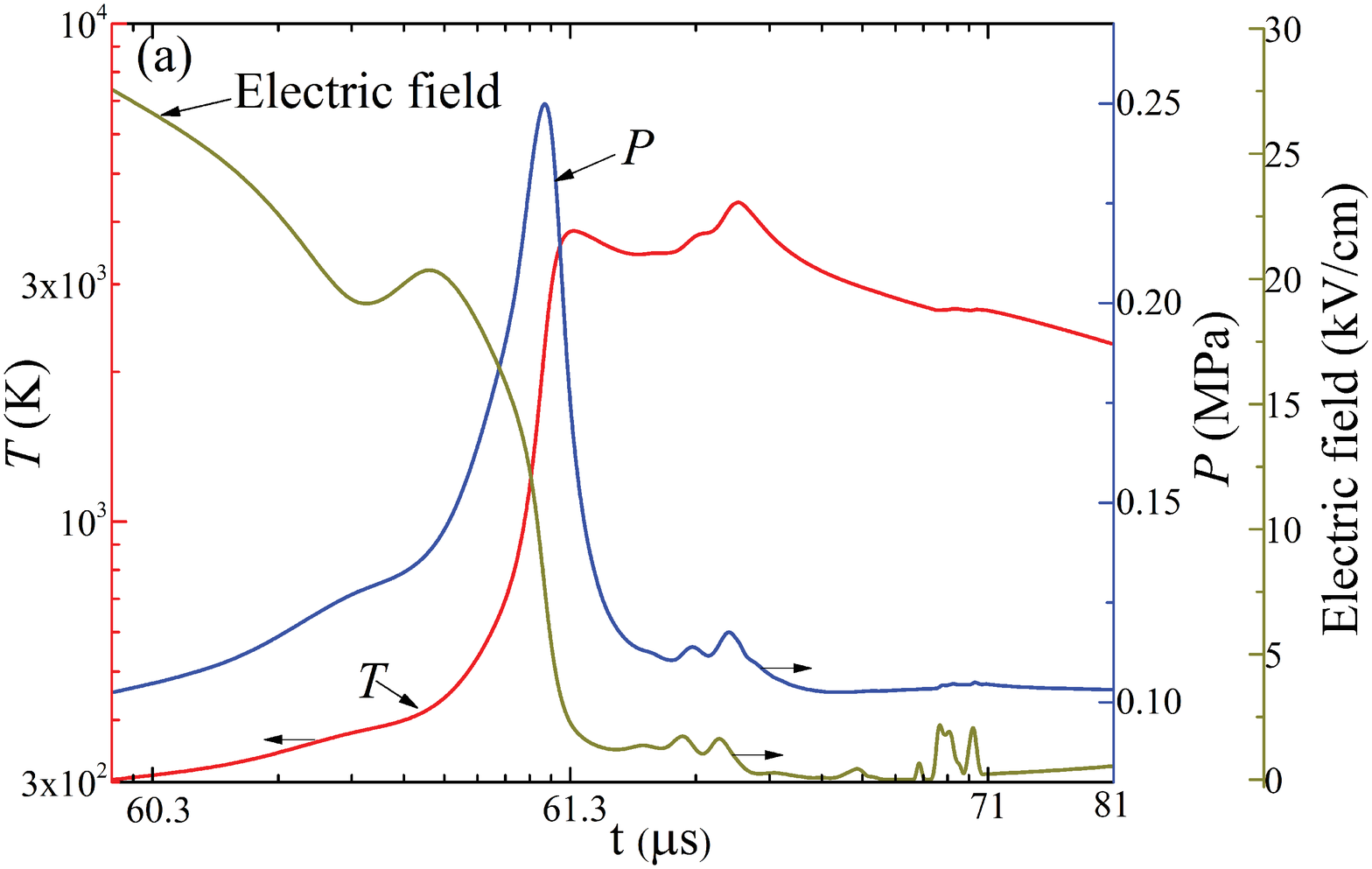}}
\subfigure[]{
\label{fig6b} \includegraphics[width=0.45\textwidth, height=0.32\textwidth]{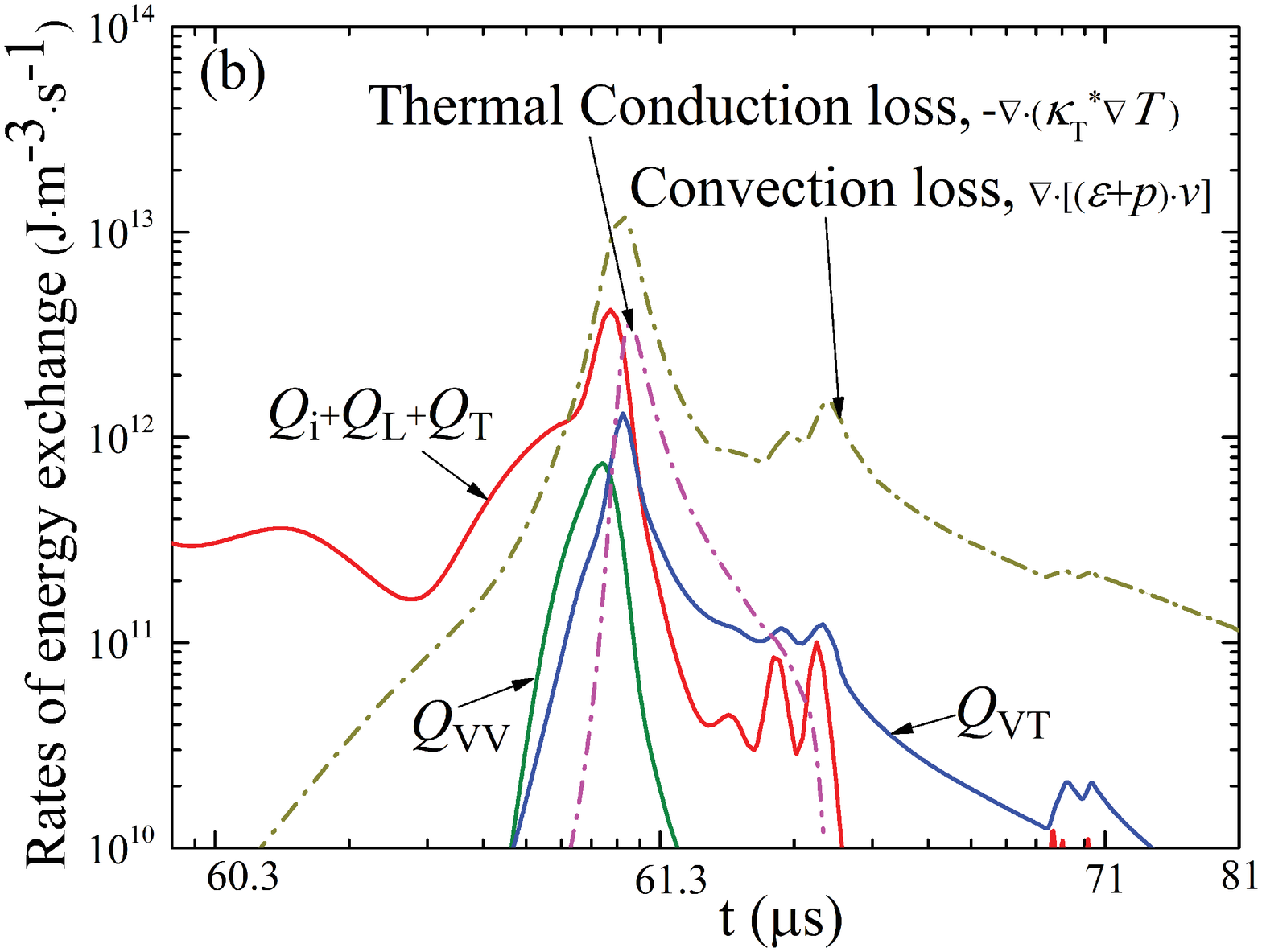}}
\caption{Temporal dynamics of (a) temperature $T$, pressure $P$, electric field, and (b) rates of energy exchange at the discharge axis (r=0), obtained from the model. The times are consistent with those in FIG. 2. In (b), the effective injected translational energy (solid lines) includes the following: $Q_{\text{i}}$ is Joule heating from the ion current, $Q_{\text{L}}$ is from elastic collisions, $Q_{\text{T}}$ is from the quenching of excited electronic states, $Q_{\text{VV}}$ is relaxation from upper levels, and $Q_{\text{VT}}$ is vibrational-translational relaxation of the first excited vibrational level in N$_2$. Translational energy loss (dotted lines) includes convection and thermal conduction loss.
}
\label{fig6}
\end{figure*}

In summary, the continuous temporal evolution of the plasma channel and the shock wave was recorded by a Mach-Zehnder interferometer in combination with a high-speed video camera, and reproduced by a thermo-hydrodynamic model with the experiment measured current as its input. The simulation results for the shock wave front positions and the radius of the plasma channel showed good consistency with the experimental measurements. The detailed thermal parameters obtained by the simulation showed that continuous energy injection by the current results in a temporary over-pressure process in the plasma channel and produces the shock wave.

\section*{Supplementary Material}
See the supplementary material for the evolution of a shock wave.

\section*{Acknowledgements}
This work was supported by the National Science Foundation of China under grants 51577098 and 51325703.


%
%

%



\begin{thebibliography}{14}%
\makeatletter
\providecommand \@ifxundefined [1]{%
 \@ifx{#1\undefined}
}%
\providecommand \@ifnum [1]{%
 \ifnum #1\expandafter \@firstoftwo
 \else \expandafter \@secondoftwo
 \fi
}%
\providecommand \@ifx [1]{%
 \ifx #1\expandafter \@firstoftwo
 \else \expandafter \@secondoftwo
 \fi
}%
\providecommand \natexlab [1]{#1}%
\providecommand \enquote  [1]{``#1''}%
\providecommand \bibnamefont  [1]{#1}%
\providecommand \bibfnamefont [1]{#1}%
\providecommand \citenamefont [1]{#1}%
\providecommand \href@noop [0]{\@secondoftwo}%
\providecommand \href [0]{\begingroup \@sanitize@url \@href}%
\providecommand \@href[1]{\@@startlink{#1}\@@href}%
\providecommand \@@href[1]{\endgroup#1\@@endlink}%
\providecommand \@sanitize@url [0]{\catcode `\\12\catcode `\$12\catcode
  `\&12\catcode `\#12\catcode `\^12\catcode `\_12\catcode `\%12\relax}%
\providecommand \@@startlink[1]{}%
\providecommand \@@endlink[0]{}%
\providecommand \url  [0]{\begingroup\@sanitize@url \@url }%
\providecommand \@url [1]{\endgroup\@href {#1}{\urlprefix }}%
\providecommand \urlprefix  [0]{URL }%
\providecommand \Eprint [0]{\href }%
\providecommand \doibase [0]{http://dx.doi.org/}%
\providecommand \selectlanguage [0]{\@gobble}%
\providecommand \bibinfo  [0]{\@secondoftwo}%
\providecommand \bibfield  [0]{\@secondoftwo}%
\providecommand \translation [1]{[#1]}%
\providecommand \BibitemOpen [0]{}%
\providecommand \bibitemStop [0]{}%
\providecommand \bibitemNoStop [0]{.\EOS\space}%
\providecommand \EOS [0]{\spacefactor3000\relax}%
\providecommand \BibitemShut  [1]{\csname bibitem#1\endcsname}%
\let\auto@bib@innerbib\@empty
\bibitem [{\citenamefont {da~Silva}\ \emph {et~al.}(2017)\citenamefont
  {da~Silva}, \citenamefont {Millan}, \citenamefont {McGaw}, \citenamefont
  {Yu}, \citenamefont {Putter}, \citenamefont {LaBelle},\ and\ \citenamefont
  {Dwyer}}]{silva2017}%
  \BibitemOpen
  \bibfield  {author} {\bibinfo {author} {\bibfnamefont {C.~L.}\ \bibnamefont
  {da~Silva}}, \bibinfo {author} {\bibfnamefont {R.~M.}\ \bibnamefont
  {Millan}}, \bibinfo {author} {\bibfnamefont {D.~G.}\ \bibnamefont {McGaw}},
  \bibinfo {author} {\bibfnamefont {C.~T.}\ \bibnamefont {Yu}}, \bibinfo
  {author} {\bibfnamefont {A.~S.}\ \bibnamefont {Putter}}, \bibinfo {author}
  {\bibfnamefont {J.}~\bibnamefont {LaBelle}}, \ and\ \bibinfo {author}
  {\bibfnamefont {J.}~\bibnamefont {Dwyer}},\ }\href@noop {} {\bibfield
  {journal} {\bibinfo  {journal} {Geophys. Res. Lett.}\ }\textbf {\bibinfo
  {volume} {44}},\ \bibinfo {pages} {11174} (\bibinfo {year}
  {2017})}\BibitemShut {NoStop}%
\bibitem [{\citenamefont {Biggerstaff}\ \emph {et~al.}(2017)\citenamefont
  {Biggerstaff}, \citenamefont {Zounes}, \citenamefont {Alford}, \citenamefont
  {Carrie}, \citenamefont {Pilkey}, \citenamefont {Uman},\ and\ \citenamefont
  {Jordan}}]{big2017}%
  \BibitemOpen
  \bibfield  {author} {\bibinfo {author} {\bibfnamefont {M.~I.}\ \bibnamefont
  {Biggerstaff}}, \bibinfo {author} {\bibfnamefont {Z.}~\bibnamefont {Zounes}},
  \bibinfo {author} {\bibfnamefont {A.}~\bibnamefont {Alford}}, \bibinfo
  {author} {\bibfnamefont {G.}~\bibnamefont {Carrie}}, \bibinfo {author}
  {\bibfnamefont {J.~T.}\ \bibnamefont {Pilkey}}, \bibinfo {author}
  {\bibfnamefont {M.}~\bibnamefont {Uman}}, \ and\ \bibinfo {author}
  {\bibfnamefont {D.}~\bibnamefont {Jordan}},\ }\href@noop {} {\bibfield
  {journal} {\bibinfo  {journal} {Geophys. Res. Lett.}\ }\textbf {\bibinfo
  {volume} {44}},\ \bibinfo {pages} {8027} (\bibinfo {year}
  {2017})}\BibitemShut {NoStop}%
\bibitem [{\citenamefont {Zhang}\ \emph {et~al.}(2017)\citenamefont {Zhang},
  \citenamefont {Zhu}, \citenamefont {Gu},\ and\ \citenamefont {He}}]{zhu2017}%
  \BibitemOpen
  \bibfield  {author} {\bibinfo {author} {\bibfnamefont {X.}~\bibnamefont
  {Zhang}}, \bibinfo {author} {\bibfnamefont {Y.}~\bibnamefont {Zhu}}, \bibinfo
  {author} {\bibfnamefont {S.}~\bibnamefont {Gu}}, \ and\ \bibinfo {author}
  {\bibfnamefont {J.}~\bibnamefont {He}},\ }\href@noop {} {\bibfield  {journal}
  {\bibinfo  {journal} {Appl. Phys. Lett.}\ }\textbf {\bibinfo {volume}
  {111}},\ \bibinfo {pages} {224101} (\bibinfo {year} {2017})}\BibitemShut
  {NoStop}%
\bibitem [{\citenamefont {Razzaghi}\ \emph {et~al.}(2018)\citenamefont
  {Razzaghi}, \citenamefont {Scatena}, \citenamefont {Sheshyekani},
  \citenamefont {Paolone}, \citenamefont {Rachidi},\ and\ \citenamefont
  {Antonini}}]{far2018}%
  \BibitemOpen
  \bibfield  {author} {\bibinfo {author} {\bibfnamefont {R.}~\bibnamefont
  {Razzaghi}}, \bibinfo {author} {\bibfnamefont {M.}~\bibnamefont {Scatena}},
  \bibinfo {author} {\bibfnamefont {K.}~\bibnamefont {Sheshyekani}}, \bibinfo
  {author} {\bibfnamefont {M.}~\bibnamefont {Paolone}}, \bibinfo {author}
  {\bibfnamefont {F.}~\bibnamefont {Rachidi}}, \ and\ \bibinfo {author}
  {\bibfnamefont {G.}~\bibnamefont {Antonini}},\ }\href@noop {} {\bibfield
  {journal} {\bibinfo  {journal} {Electr. Power Syst. Res.}\ }\textbf {\bibinfo
  {volume} {160}},\ \bibinfo {pages} {282} (\bibinfo {year}
  {2018})}\BibitemShut {NoStop}%
\bibitem [{\citenamefont {Thang}\ \emph {et~al.}(2018)\citenamefont {Thang},
  \citenamefont {Baba}, \citenamefont {Itamoto},\ and\ \citenamefont
  {Rakov}}]{rakov2018}%
  \BibitemOpen
  \bibfield  {author} {\bibinfo {author} {\bibfnamefont {T.~H.}\ \bibnamefont
  {Thang}}, \bibinfo {author} {\bibfnamefont {Y.}~\bibnamefont {Baba}},
  \bibinfo {author} {\bibfnamefont {N.}~\bibnamefont {Itamoto}}, \ and\
  \bibinfo {author} {\bibfnamefont {V.~A.}\ \bibnamefont {Rakov}},\ }\href@noop
  {} {\bibfield  {journal} {\bibinfo  {journal} {Electr. Power Syst. Res.}\
  }\textbf {\bibinfo {volume} {159}},\ \bibinfo {pages} {17} (\bibinfo {year}
  {2018})}\BibitemShut {NoStop}%
\bibitem [{\citenamefont {Zhou}\ \emph {et~al.}(2015)\citenamefont {Zhou},
  \citenamefont {Zeng}, \citenamefont {Zhuang},\ and\ \citenamefont
  {Chen}}]{pop2015}%
  \BibitemOpen
  \bibfield  {author} {\bibinfo {author} {\bibfnamefont {X.}~\bibnamefont
  {Zhou}}, \bibinfo {author} {\bibfnamefont {R.}~\bibnamefont {Zeng}}, \bibinfo
  {author} {\bibfnamefont {C.}~\bibnamefont {Zhuang}}, \ and\ \bibinfo {author}
  {\bibfnamefont {S.}~\bibnamefont {Chen}},\ }\href@noop {} {\bibfield
  {journal} {\bibinfo  {journal} {Phys. Plasmas}\ }\textbf {\bibinfo {volume}
  {22}},\ \bibinfo {pages} {063508} (\bibinfo {year} {2015})}\BibitemShut
  {NoStop}%
\bibitem [{\citenamefont {Suzuki}(1971)}]{suzuki1971}%
  \BibitemOpen
  \bibfield  {author} {\bibinfo {author} {\bibfnamefont {T.}~\bibnamefont
  {Suzuki}},\ }\href@noop {} {\bibfield  {journal} {\bibinfo  {journal} {J.
  Phys. D: Appl. Phys.}\ }\textbf {\bibinfo {volume} {42}},\ \bibinfo {pages}
  {3766} (\bibinfo {year} {1971})}\BibitemShut {NoStop}%
\bibitem [{\citenamefont {Marode}(1975)}]{marode1975}%
  \BibitemOpen
  \bibfield  {author} {\bibinfo {author} {\bibfnamefont {E.}~\bibnamefont
  {Marode}},\ }\href@noop {} {\bibfield  {journal} {\bibinfo  {journal} {J.
  Appl. Phys.}\ }\textbf {\bibinfo {volume} {46}},\ \bibinfo {pages} {2005}
  (\bibinfo {year} {1975})}\BibitemShut {NoStop}%
\bibitem [{\citenamefont {Popov}(2003)}]{popov2003}%
  \BibitemOpen
  \bibfield  {author} {\bibinfo {author} {\bibfnamefont {N.~A.}\ \bibnamefont
  {Popov}},\ }\href@noop {} {\bibfield  {journal} {\bibinfo  {journal} {Plasma
  Phys. Rep.}\ }\textbf {\bibinfo {volume} {29}},\ \bibinfo {pages} {695}
  (\bibinfo {year} {2003})}\BibitemShut {NoStop}%
\bibitem [{\citenamefont {da~Silva}\ and\ \citenamefont
  {Pasko}(2013)}]{pasko2013}%
  \BibitemOpen
  \bibfield  {author} {\bibinfo {author} {\bibfnamefont {C.~L.}\ \bibnamefont
  {da~Silva}}\ and\ \bibinfo {author} {\bibfnamefont {V.}~\bibnamefont
  {Pasko}},\ }\href@noop {} {\bibfield  {journal} {\bibinfo  {journal} {J.
  Geophys. Res. Atmos.}\ }\textbf {\bibinfo {volume} {118}},\ \bibinfo {pages}
  {13} (\bibinfo {year} {2013})}\BibitemShut {NoStop}%
\bibitem [{\citenamefont {Shneider}, \citenamefont {Zheltikov},\ and\
  \citenamefont {Miles}(2011)}]{mikal2011}%
  \BibitemOpen
  \bibfield  {author} {\bibinfo {author} {\bibfnamefont {M.~N.}\ \bibnamefont
  {Shneider}}, \bibinfo {author} {\bibfnamefont {A.~M.}\ \bibnamefont
  {Zheltikov}}, \ and\ \bibinfo {author} {\bibfnamefont {R.~B.}\ \bibnamefont
  {Miles}},\ }\href@noop {} {\bibfield  {journal} {\bibinfo  {journal} {Phys.
  Plasmas.}\ }\textbf {\bibinfo {volume} {18}},\ \bibinfo {pages} {063509}
  (\bibinfo {year} {2011})}\BibitemShut {NoStop}%
\bibitem [{\citenamefont {Kartashov}\ and\ \citenamefont
  {Shneider}(2017)}]{mikal2017}%
  \BibitemOpen
  \bibfield  {author} {\bibinfo {author} {\bibfnamefont {D.}~\bibnamefont
  {Kartashov}}\ and\ \bibinfo {author} {\bibfnamefont {M.~N.}\ \bibnamefont
  {Shneider}},\ }\href@noop {} {\bibfield  {journal} {\bibinfo  {journal} {J.
  Appl. Phys.}\ }\textbf {\bibinfo {volume} {121}},\ \bibinfo {pages} {113303}
  (\bibinfo {year} {2017})}\BibitemShut {NoStop}%
\bibitem [{\citenamefont {Gottlieb}, \citenamefont {Shu},\ and\ \citenamefont
  {Tadmor}(2001)}]{siam2001}%
  \BibitemOpen
  \bibfield  {author} {\bibinfo {author} {\bibfnamefont {S.}~\bibnamefont
  {Gottlieb}}, \bibinfo {author} {\bibfnamefont {C.}~\bibnamefont {Shu}}, \
  and\ \bibinfo {author} {\bibfnamefont {E.}~\bibnamefont {Tadmor}},\
  }\href@noop {} {\bibfield  {journal} {\bibinfo  {journal} {SIAM Rev.}\
  }\textbf {\bibinfo {volume} {43}},\ \bibinfo {pages} {89} (\bibinfo {year}
  {2001})}\BibitemShut {NoStop}%
\bibitem [{\citenamefont {Zhou}\ \emph {et~al.}(2016)\citenamefont {Zhou},
  \citenamefont {Zeng}, \citenamefont {Li},\ and\ \citenamefont
  {Zhuang}}]{epsr2016}%
  \BibitemOpen
  \bibfield  {author} {\bibinfo {author} {\bibfnamefont {X.}~\bibnamefont
  {Zhou}}, \bibinfo {author} {\bibfnamefont {R.}~\bibnamefont {Zeng}}, \bibinfo
  {author} {\bibfnamefont {Z.}~\bibnamefont {Li}}, \ and\ \bibinfo {author}
  {\bibfnamefont {C.}~\bibnamefont {Zhuang}},\ }\href@noop {} {\bibfield
  {journal} {\bibinfo  {journal} {Electr. Power Syst. Res.}\ }\textbf {\bibinfo
  {volume} {139}} (\bibinfo {year} {2016})}\BibitemShut {NoStop}%
\end{thebibliography}

%

\end{document}